\documentclass[aps,prb,twocolumn,superscriptaddress,showpacs]{revtex4}

\usepackage[T1]{fontenc}
\usepackage{amssymb}
\usepackage{amsbsy}
\usepackage{amsmath}           
\usepackage{epsfig}

\begin{document}


\def\beq{\begin{equation}}
\def\eeq{\end{equation}}
\def\bleq{\begin{eqnarray}}
\def\eleq{\end{eqnarray}}

\newcommand{\Tr}{{\rm Tr}} 
\newcommand{\tr}{{\rm tr}} 
\newcommand{\sgn}{{\rm sgn}} 
\newcommand{\mean}[1]{\langle #1 \rangle}
\newcommand{\commu}[2]{[#1,#2]} 
\newcommand{\ket}[1]{|#1\rangle}
\newcommand{\bra}[1]{\langle#1|}
\newcommand{\braket}[2]{\langle #1|#2\rangle}
\newcommand{\dbraket}[3]{\langle #1|#2|#3\rangle} 
\newcommand{\vac}{|0\rangle} 
\newcommand{\const}{{\rm const}}

\newcommand{\jhatbf}{\hat {\textbf \j}} 
\newcommand{\Jhatbf}{\hat {\textbf \J}} 
\newcommand{\jhat}{\hat {\jmath}} 
\newcommand{\Jhat}{\hat {J}} 
\newcommand{\jbf}{\textbf j}
\newcommand{\Jbf}{\textbf J}

\def\down{\downarrow}
\def\eps{\epsilon}
\def\gam{\gamma} 
\def\Ome{\Omega}
\def\bfOme{\boldsymbol{\Omega}} 
\def\sig{\sigma}
\def\bfsig{\boldsymbol{\sigma}} 
\def\The{\Theta} 
\def\up{\uparrow}

\def\xik{\xi_{\bf k}} 
\def\Ek{E_{\bf k}}

\def\half{\frac{1}{2}}
\def\quarter{\frac{1}{4}}

\def\a{{\bf a}}
\def\b{{\bf b}}
\def\k{{\bf k}}
\def\n{{\bf n}} 
\def\p{{\bf p}} 
\def\q{{\bf q}}
\def\r{{\bf r}}
\def\t{{\bf t}}
\def\v{{\bf v}}
\def\x{{\bf x}}
\def\z{{\bf z}} 
\def\A{{\bf A}}
\def\B{{\bf B}}
\def\D{{\bf D}} 
\def\E{{\bf E}} 
\def\F{{\bf F}} 
\def\H{{\bf H}}  
\def\J{{\bf J}}
\def\K{{\bf K}} 
\def\L{{\bf L}}
\def\M{{\bf M}}  
\def\O{{\bf O}} 
\def\P{{\bf P}} 
\def\Q{{\bf Q}} 
\def\R{{\bf R}}
\def\S{{\bf S}}
\def\bfnabla{\boldsymbol{\nabla}}
\def\bfsigma{\boldsymbol{\sigma}} 
\def\nablabf{\boldsymbol{\nabla}}
\def\sigmabf{\boldsymbol{\sigma}} 

\def\para{\parallel}
\def\kpara{{k_\parallel}}
\def\kperp{{k_\perp}} 
\def\kperpp{{k_\perp'}} 
\def\qperp{{q_\perp}} 
\def\tperp{{t_\perp}} 

\def\w{\omega}
\def\wn{\omega_n}
\def\wnu{\omega_\nu}
\def\wp{\omega_p} 
\def\dmu{{\partial_\mu}} 
\def\dt{{\partial_t}}
\def\dtau{{\partial_\tau}}  
\def\det{{\rm det}} 

\def\intt{\int_{-\infty}^\infty dt\,} 
\def\inttp{\int_{-\infty}^\infty dt'\,} 
\def\intk{\int_{\bf k}} 
\def\intr{\int d^3r\,} 
\def\intrp{\int d^3r'\,}
\def\inttau{\int_0^\beta d\tau}
\def\inttaup{\int_0^\beta d\tau'}
\def\intx{\int d^4x\,} 
\def\inttaur{\int_0^\beta d\tau \int d^3r\,}
\def\intinf{\int_{-\infty}^\infty}

\def\calC{{\cal C}}
\def\calD{{\cal D}}
\def\calF{{\cal F}} 
\def\calG{{\cal G}}
\def\calH{{\cal H}}
\def\calL{{\cal L}} 
\def\calO{{\cal O}}
\def\calP{{\cal P}}  
\def\calR{{\cal R}} 
\def\calS{{\cal S}}
\def\calT{{\cal T}}
\def\calU{{\cal U}}


\title{Variational Cluster Perturbation Theory for Bose-Hubbard models } 

\author{W. Koller}
\affiliation{Department of Mathematics, Imperial College, 
180 Queen's Gate, London SW7 2AZ, UK}
\author{N. Dupuis}
\affiliation{Department of Mathematics, Imperial College, 
180 Queen's Gate, London SW7 2AZ, UK}
\affiliation{Laboratoire de Physique des Solides, CNRS UMR 8502,
  Universit\'e Paris-Sud, 91405 Orsay, France}

\date{September 11, 2006}

\begin{abstract}  
We discuss the application of the variational cluster perturbation theory
(VCPT) to the Mott-insulator--to--superfluid transition in the Bose-Hubbard
model. We show how the VCPT can be formulated in such a way that it 
gives a translation invariant excitation spectrum -- free of spurious gaps -- 
despite the fact that if formally breaks translation invariance.
The phase diagram and the single-particle Green function in the
insulating phase are obtained for one-dimensional systems. When the
chemical potential of the cluster is taken as a variational parameter, the 
VCPT reproduces the dimension dependence of the phase diagram even for one-site
clusters. We find a good quantitative
agreement with the results of the density-matrix renormalization group when 
the number of sites in the cluster becomes of order 10. The extension of the
method to the superfluid phase is discussed.   
\end{abstract} 


\maketitle

\section{Introduction}

The Bose-Hubbard model describes interacting bosons on a lattice. It provides
a generic description of the quantum phase transition between superfluid
(SF) and Mott-insulator (MI) states observed in condensed-matter systems such
as Josephson junction arrays or granular
superconductors,\cite{Fisher89} as well as in ultracold atoms in optical
lattices.\cite{Jaksch98,Greiner02,Stoferle04,Gerbier05} The remarkable degree
of experimental control over all the relevant parameters (density, interaction
strength, lattice geometry and dimensionality) in ultracold atoms makes
possible a detailed study of the MI-SF transition. 

The Bose-Hubbard model has been studied numerically using the Gutzwiller
mean-field ansatz,\cite{Rokhsar91,Krauth92,Schroll04} the density-matrix
renormalization group,\cite{Kuhner98,Kuhner00} exact
diagonalizations,\cite{Roth03a,Roth03b} and 
quantum Monte Carlo.\cite{Batrouni90,Batrouni92,Krauth91} (More recent works
include the harmonic trap potential that confines the ultracold atomic gases;
see for instance Ref.~\onlinecite{Wessel04}.) 
Most analytical approaches rely on a perturbation theory that
assumes the kinetic energy to be small and treats exactly the on-site
repulsion. The intersite hopping is taken into account either at the
mean-field level or in perturbation in a strong-coupling
expansion.\cite{Fisher89,Sheshadri93,Kampf93,VanOosten01,Sachdev99,Dickerscheid03,Konabe04,Sengupta05,Freericks94,Freericks96,Elstner99,Buonsante05}

In this paper, we apply the variational cluster perturbation theory (VCPT) to
the Bose-Hubbard model at commensurate density. The VCPT has been developed for
strongly-correlated fermion systems. It is an extension of cluster
perturbation theory (CPT) that is based on the self-energy functional approach
(SFA). Within the CPT,\cite{Gros93,Senechal00,Senechal02} the 
lattice is partitioned into disconnected identical clusters. The Hamiltonian
of the cluster is solved numerically, and the intercluster hopping is then
treated perturbatively to leading order in a strong-coupling expansion. 
Contrary to exact diagonalizations of small systems, the CPT provides results
in the thermodynamic limit, and the single-particle Green function is defined 
for any wave vector in the Brillouin zone. When based on a single-site
cluster, the CPT yields the Hubbard I approximation;\cite{Hubbard63} applied
to bosonic systems, it reproduces the leading order of the aforementioned
strong-coupling theory.\cite{Sengupta05} 

The SFA
is based on the variational principle $\delta\Omega[\Sigma]/\delta\Sigma=0$
for the grand potential expressed as a functional of the self-energy
$\Sigma$.\cite{Potthoff03a,Potthoff03b,Potthoff03c} The stationary condition
is solved within a restricted space of self-energies taken from a reference
system that can be solved numerically. Within the
VCPT,\cite{Potthoff03c,Dahnken04} the reference system consists of
disconnected identical clusters. The VCPT improves on the CPT since 
the parameters of the intracluster kinetic Hamiltonian are variational. In
particular, this enables to consider broken-symmetry states. The VCPT has been
used with some success to study strongly correlated electron
systems.\cite{Dahnken04,Aichhorn04,Senechal05,Aichhorn05,Danhken05} 

The outline of the paper is as follows. In Sec.~\ref{sec_vcpt}, we describe
the SFA and the VCPT for bosonic models in the absence of superfluidity. We
modify the original formulation of the SFA\cite{Potthoff03a} to ensure that 
final results remain translation invariant regardless of the choice of the
reference system. We also stress the necessity to consider the chemical
potential of the cluster as a variational parameter. In Sec.~\ref{sec_qunu},
we present numerical results for the phase diagram and the single-particle
Green function in a one-dimensional (1D) system.
Even in the simplest case of a reference system consisting of 
single-site clusters, the VCPT improves drastically on the CPT. In particular,
we obtain the correct form of the Mott lobes in the
($t/U,\mu/U$) phase diagram ($t$ is the hopping amplitude, $U$ the onsite
repulsion, and $\mu$ the chemical 
potential). We find a good quantitative agreement with the results of the
density-matrix renormalization group\cite{Kuhner00} when  
the number of sites in the cluster becomes of order 10. The
last section is devoted to a summary of our results and a discussion of future
developments, in particular the extension of the VCPT to the superfluid phase.

\section{Variational Cluster Perturbation Theory for bosons}
\label{sec_vcpt}

The Bose-Hubbard model is defined by the Hamiltonian
\beq
H = - \sum_{\r,\r'} (\psi^\dagger_{\bf r}
t_{\r\r'} \psi_{{\bf r}'} + {\rm H.c.}) - \mu \sum_{\bf r} n_{\bf r} +
\frac{U}{2} \sum_{\bf r} n_{\bf r}(n_{\bf r}-1) ,
\label{ham}
\eeq 
where $\psi_{\bf r},\psi^\dagger_{\bf r}$ are annihilation/creation bosonic
operators and $n_{\bf r}=\psi^\dagger_{\bf r}\psi_{\bf r}$. The discrete
variable ${\bf r}$ labels the sites of the lattice, which is assumed to be
bipartite with coordination number $z$.  
The hopping matrix $\hat t$ satisfies $t_{\r\r'}=t>0$ if $\r$ and
$\r'$ are nearest neighbors and vanishes otherwise; this assumption can
however be easily relaxed and longer-range hopping considered. 
For reasons explained below, the boson-boson interaction should be onsite.
The density $n$, i.e. the average number of bosons per site, is fixed by
the chemical potential $\mu$. In the following, we shall consider only the
Mott phase and the zero-temperature limit. 

\subsection{General formalism}

In the absence of superfluidity ($\mean{\psi_{\r}}=0$), the grand
potential (per lattice site) $\Omega$ and the single-particle Green function 
$G$ can be obtained from the stationary point 
of the functional\cite{Luttinger60,Dedominicis64a,Dedominicis64b}
\beq
\Omega[G] = \frac{1}{N\beta} \left\lbrace \Tr\ln(-G^{-1}) + \Tr(G_0^{-1}G-1) +
\Phi[G] \right\rbrace ,
\eeq 
where $G_0$ is the non-interacting Green function and $\Phi[G]$ the
Luttinger-Ward functional. $\beta=1/T$ is the inverse temperature and $N$ the
number of lattice sites. $\Tr$ denotes a trace over space and time
indices. The stationary condition $\delta\Omega[G]/\delta G=0$ yields the
Dyson equation  
\beq
G^{-1} = G_0^{-1} - \Sigma ,
\label{Dyson}
\eeq
where the self-energy $\Sigma$ is defined by
\beq
\Sigma_{ij} = - \frac{\delta \Phi[G]}{\delta G_{ji}} .
\label{Sigma_def}
\eeq 
$i$ and $j$ are collective indices that label position and time.

The SFA is based on the functional\cite{Potthoff03a}
\beq
\Omega[\Sigma] = \frac{1}{N\beta} \left\lbrace \Tr\ln(-G_0^{-1}+\Sigma) +
F[\Sigma] \right\rbrace ,
\label{Omega0}
\eeq
where 
\beq 
F[\Sigma] = \Phi[G] + \Tr(\Sigma G) 
\label{F_def} 
\eeq
is the Legendre transform of $\Phi[G]$. In
Eq.~(\ref{F_def}), $G$ should be considered as a functional $G[\Sigma]$ of the
self-energy obtained by inverting (\ref{Sigma_def}). $F$ satisfies
\beq
\frac{\delta F[\Sigma]}{\delta\Sigma_{ij}} = G_{ji} , 
\eeq 
and the stationary condition $\delta\Omega[\Sigma]/\delta\Sigma=0$
reproduces the Dyson equation (\ref{Dyson}). 

So far, we have followed the approach of Ref.~\onlinecite{Potthoff03a} with
minor modifications due to the fact that we consider bosons instead of
fermions. We now introduce a refinement of the approach, the motivation of
which will be discussed below. In a translation invariant system,
the actual self-energy is diagonal in $\k$ space: $\Sigma(\k,\k',z) =
\delta_{\k,\k'} \Sigma(\k,\k,z)$ ($z$ is a complex frequency). Without
changing its value at the stationary point, we can therefore modify the
functional (\ref{Omega0}) into 
\beq 
\Omega[\Sigma] = \frac{1}{N\beta} \bigl\lbrace \Tr\ln(-G_0^{-1}+\tilde\Sigma) +
F[\Sigma] \bigr\rbrace ,
\label{Omega_new}
\eeq
where
\beq
\tilde\Sigma(\k,\k',z) = \delta_{\k,\k'} \Sigma(\k,\k,z) 
\eeq
is the diagonal part of the self-energy. One easily verifies that the
stationary condition $\delta\Omega[\Sigma]/\delta\Sigma=0$ yields the Dyson
equation 
\beq
G^{-1}(\k,z) = G_0^{-1}(\k,z) - \tilde\Sigma(\k,z) ,
\label{Sigma_tilde}
\eeq 
where $G_0^{-1}(\k,i\wn)=i\wn+\mu-\eps_\k$. $\eps_\k$ is the Fourier transform
of $-t_{\r,\r'}$ and gives the lattice dispersion of the bosons.
In Eq.~(\ref{Sigma_tilde}), $\tilde\Sigma(\k,z)$ denotes
$\tilde\Sigma(\k,\k,z)$, etc. Thus the two functionals (\ref{Omega0}) and
(\ref{Omega_new}) contain the same information in the case of translation
invariant systems. 

Let us now consider the two Hamiltonians\cite{Potthoff03a} 
\bleq
H(x) &=& H_0(x) + H_U , \nonumber \\ 
H(x') &=& H_0(x') + H_U .
\eleq 
$H(x)$ is the Hamiltonian of the Bose-Hubbard model [Eq.~(\ref{ham})], and
$H(x')$ that of a reference system. Both Hamiltonians are defined on the same
lattice and share the same interaction Hamiltonian $H_U$.  
The kinetic parts $H_0(x)$ and $H_0(x')$ include a chemical potential
term. $x$ stands for the parameters on which $H_0$ depends: the intersite
hopping matrix $\hat t$ and the chemical potential $\mu$.
The Hamiltonian $H(x)$ is translation invariant, but that of the reference
system, $H(x')$, may not be. The Luttinger-Ward functional $\Phi[G]$ is given
by the sum of the two-particle irreducible (skeleton)
diagrams\cite{Dedominicis64a,Dedominicis64b} and is independent of $G_0$. 
It follows that $H(x)$ and $H(x')$ share the same Luttinger-Ward
functional $\Phi[G]$ and therefore the same functional
$F[\Sigma]$\cite{Potthoff03a}. This leads us to consider the functionals
\bleq
\Omega_x[\Sigma] &=&  \frac{1}{N\beta} \bigl\lbrace
\Tr\ln(-G_0^{-1}+\tilde\Sigma) + F[\Sigma] \bigr\rbrace , \nonumber \\ 
\Omega_{x'}[\Sigma] &=&  \frac{1}{N\beta} \bigl\lbrace
\Tr\ln(-G_0'{}^{-1}+\Sigma) + F[\Sigma] \bigr\rbrace ,
\label{Omega1}
\eleq 
where $G_0\equiv G_0(x)$ and $G_0'\equiv G_0(x')$ are the Green functions
corresponding to $H_0(x)$ and  $H_0(x')$, respectively. 
Note that we have taken advantage of the translation invariance of $H(x)$ in
writing $\Omega_x[\Sigma]$. The unknown functional $F[\Sigma]$ can now be
eliminated by taking the difference of the two preceeding equations,
\begin{multline}
\Omega_x[\Sigma] = \Omega_{x'}[\Sigma]  + \frac{1}{N\beta} \bigl\lbrace
\Tr\ln(-G_0^{-1}+\tilde\Sigma) \\ - \Tr\ln(-G_0'{}^{-1}+\Sigma) \bigr\rbrace .
\label{Omega1a}
\end{multline} 

The form (\ref{Omega1a}) of the self-energy functional $\Omega_x[\Sigma]$ is
exact. To make the determination of the stationary points of
$\Omega_x[\Sigma]$ possible, one has to restrict the space of self-energies. A
natural approximation is to evaluate $\Omega_x[\Sigma]$ for the (physical)
self-energies $\Sigma(x')$ obtained from the reference system -- assuming that
the reference Hamiltonian can be solved exactly.\cite{Potthoff03a} The
functional $\Omega_x[\Sigma(x')]\equiv\Omega_x(x')$ becomes a function of
$x'$,  
\begin{multline}
\Omega_x(x') = \Omega' + \frac{1}{N\beta} \bigl\lbrace
\Tr\ln[-G_0^{-1}+\tilde\Sigma(x')] \\  
- \Tr\ln[-G'_0{}^{-1}+\Sigma(x')] \bigr\rbrace
\label{Omega2}
\end{multline}
where $\Omega'=\Omega_{x'}[\Sigma(x')]$ is the exact grand potential of the
reference system. The stationary condition becomes 
\beq
\frac{\partial\Omega_x(x')}{\partial x'} = 0 .
\label{station}
\eeq 
Note that the stationary point does not have to be a minimum. We shall see
that it is actually never a minimum when the chemical potential of the
reference system is taken as a variational parameter. This conclusion also
holds in fermion systems.\cite{Aichhorn05}

\subsection{Reference system} 

In the VCPT one considers a reference system consisting of a superlattice 
of $N_c=N/L$ identical clusters with no intercluster kinetic coupling 
($L$ is the number of sites in a cluster).  
The determination of $\Omega'$ and the self-energy $\Sigma(x')$ then
requires to solve a finite size system Hamiltonian provided that the
interaction is local. For moderate values of $L$, this can be done
numerically. Eq.~(\ref{Omega2}) can be rewritten as 
\begin{multline}
\Omega_x(x') = \Omega'  - \frac{1}{N\beta} \sum_{\k,\wn} 
\ln[-G(\k,i\wn)] e^{i\wn \eta} \\
+ \frac{1}{N\beta} \sum_{\wn} \tr\ln[-G'(i\wn)] e^{i\wn \eta}
\label{Omega3} 
\end{multline}
($\wn$ is a bosonic Matsubara frequency),
where $G^{-1}=G_0^{-1}-\tilde\Sigma$ and $G'{}^{-1}=G_0'{}^{-1}-\Sigma$ (from
now on, we do not write explicitely the $x'$ dependence of $\Sigma(x')$).
$\tr$ denotes the trace over space indices. $\tr\,\ln(-G')$ can be evaluated
by considering a single cluster and multiplying the result by $N_c=N/L$ to
take into account the total number of clusters. When the
chemical potentials $\mu$ and $\mu'$ differ, the usual factor $e^{i\wn\eta}$
($\eta\to 0^+$) is necessary for the sum over $\wn$ to converge. 

Let us now discuss the motivation for using the functional (\ref{Omega_new})
rather than (\ref{Omega0}). 
The reference system has the periodicity of the cluster superlattice, but not
that of the translation invariant system of interest. Except for single-site
clusters, the self-energy $\Sigma$ is not diagonal in $\k$ space. By making
use of the functional (\ref{Omega_new}), we ensure the translation invariance
of the Green function defined by Eq.~(\ref{Sigma_tilde}). In 
applications of the VCPT to fermionic systems, the translation invariant Green
function of the system was approximated by the diagonal part of
$G=(G_0^{-1}-\Sigma)^{-1}$. However, because $\Sigma$ itself is not translation
invariant, the resulting excitation spectrum exhibits spurious gaps that
reflect the periodicity of the reference system. To some extent, this
difficulty can be eliminated by introducing an artificial broadening of the
energy states; technically this is achieved by choosing a sufficiently large
value of $\eta$. In bosonic systems, even when one allows for an energy
broadening, these gaps are so pronounced that the excitation spectrum does not
bear much physical meaning. 

The translation invariant self-energy $\tilde\Sigma(\k,z)$ turns out to be
simply related to the cluster self-energy when periodic boundary conditions --
as allowed in the VCPT, since this simply assumes a particular choice of the
(in principle variational) intersite hopping matrix $\hat t'$ -- are chosen
for the cluster. Let us write the self-energy of the reference system as 
\beq
\Sigma(\R+\r_a,\R'+\r_b,z) = \delta_{\R,\R'} \Sigma_{ab}(z) ,
\eeq
where $\Sigma_{ab}(z)$ is the cluster self-energy obtained by solving
numerically the Hamiltonian $H(x')$. 
We use the notation $\R+\r_a$ for a site of a lattice, $\R$ for a
site of the superlattice, and $\r_a$ ($a\in [1,L]$) for the position within the
cluster. The translation invariant self-energy then reads
\beq
\tilde\Sigma(\k,z) = \frac{1}{L} \sum_{a,b=1}^L e^{-i\k\cdot(\r_a-\r_b)} 
\Sigma_{ab}(z) .
\label{Sigma_c}
\eeq 
With periodic boundary conditions for the cluster, we can define the Fourier
transform of 
$\Sigma_{ab}(z)$ for any vector $\k$ of the reciprocal superlattice. In that
case, $\tilde\Sigma(\k,z)$ defined by Eq.~(\ref{Sigma_c}) is nothing but the
interpolation of the translation invariant cluster self-energy to all vectors
$\k$ of the Brillouin zone.

\subsection{Numerical implementation}

To numerically evaluate the sum over Matsubara frequencies in
Eq.~(\ref{Omega3}), one has to get rid of the convergence factor $e^{i\wn
  \eta}$. To this end, one subtracts and adds the term 
\begin{multline}
\frac{1}{\beta N} \sum_{\K,\wn} \tr_c \ln\left[ 1 - \frac{\Delta
 h_0(\K)}{i\wn-y} \right] e^{i\wn\eta}  \\ = \frac{1}{\beta N} \sum_{\K,\alpha}
\ln\left[1-e^{-\beta \left(y+\lambda_\alpha(\K)\right)}\right] -
  \frac{1}{\beta} \ln\left(1-e^{-\beta y}\right) , 
\label{subadd}
\end{multline}
where $\tr_c$ denotes the trace over intracluster space indices. 
$\Delta h_0=h_0-h_0'$ is defined from the Green functions
$G_0^{-1}(i\wn)=i\wn-h_0$ and $G_0'{}^{-1}(i\wn)=i\wn-h_0'$. $\Delta h_0(\K)$
is the Fourier transform of $\Delta h_0$ with respect to the superlattice
of clusters,  
\beq
\Delta h_0(\K)_{ab}  = \sum_\R e^{-i\K\cdot(\R-\R')} \Delta
h_0(\R+\r_a,\R'+\r_b) . 
\eeq
$ \Delta h_0(\r,\r')$ denotes the matrix elements of $\Delta h_0$ in
real space, and $\K$ is a vector of the reduced Brillouin zone
corresponding to the superlattice. $\lambda_\alpha(\K)$ 
($\alpha\in [1,L]$) are the eigenvalues of $\Delta h_0(\K)$, and the real
number $y$ should satisfy $y>\max_{\alpha,\K}|\lambda_\alpha(\K)|$. 
Eq.~(\ref{subadd}) is derived in Appendix \ref{app_subadd}. 
Combining (\ref{Omega3}) and
(\ref{subadd}), we finally obtain
\begin{multline}
\Omega_x(x') = \Omega' - \frac{1}{N\beta} \sum_{\wn} \biggl\lbrace 
\sum_{\k}  \ln[-G(\k,i\wn)] \\ - \tr\ln[-G'(i\wn)]   
+ \sum_{\K} \tr_c \ln\left[ 1 - \frac{\Delta
  h_0(\K)}{i\wn-y} \right] \biggr\rbrace \\
+ \frac{1}{\beta N} \sum_{\K,\alpha}
\ln\left[1-e^{-\beta \left(y+\lambda_\alpha(\K)\right)}\right]
- \frac{1}{\beta} \ln\left(1-e^{-\beta y}\right) . 
\label{Omega4}
\end{multline}
The term inside the curly brakets behaves as $1/\wn^2$ when $|\wn|\to\infty$,
so that the convergence factor $e^{i\wn \eta}$ is 
not necessary anymore. Eq.~(\ref{Omega4}) is the starting point of the
numerical calculations discussed in Sec.~\ref{sec_qunu}.

\subsection{$\mu'$ as a variational parameter} 
\label{subsec_mup}

In this section, we discuss the role of the chemical potential $\mu'$ of the
cluster as a variational parameter. The boson density is obtained from 
\beq
n = - \frac{\partial \Omega_x(x')}{\partial \mu} ,
\label{density1}
\eeq 
where the value of $x'=x'(x)$ is determined from the stationary condition
(\ref{station}). Combining Eqs.~(\ref{density1}) and (\ref{station}), we obtain
\beq
n = - \frac{\partial \Omega_x(x')}{\partial \mu} \biggr|_{x'} 
  - \frac{\partial \Omega_x(x')}{\partial x'} \biggr|_x \frac{\partial
    x'}{\partial \mu} =  - \frac{\partial \Omega_x(x')}{\partial \mu}
  \biggr|_{x'}   .
\eeq 
From Eq.~(\ref{Omega3}), we then deduce 
\bleq
n &=& - \frac{\partial}{\partial\mu} \frac{1}{N\beta} \sum_{\k,\wn}
\ln[-G^{-1}(\k,i\wn)]e^{i\wn \eta} \Bigr|_{x'} 
\nonumber \\ 
&=& - \frac{1}{N\beta} \Tr(G) .
\label{density3}
\eleq
Here we have used the fact that the only dependence on $\mu$ (at fixed $x'$)
comes from $G_0(\k,i\wn)$. 
Thus the stationary condition (\ref{station}) ensures that the approach is
thermodynamically consistent: the boson density can be calculated either from
the grand potential or the single-particle Green function. 

The stationary condition also ensures that the boson density $n$ of the
system is the same as that of the reference system ($n_c$). To see this, we
make use of the following result derived in Appendix \ref{app_int}: 
\begin{multline}
 \sum_{\k,\wn} \ln[-G(\k,i\wn)] e^{i\wn \eta} = \\ 
 - \sum_{\k,\gam} \ln \left|1-e^{-\beta E_\gam(\k)}\right| 
\shoveright{ +   \sum_{\k,\gam} \ln \left|1-e^{-\beta
  Z_{\gam}(\k)}\right| , }  \\
\shoveleft{ \sum_{\wn} \tr\ln[-G'(i\wn)] e^{i\wn
    \eta} = } \\ 
 - \sum_{\alpha,\gam} \ln \left|1-e^{-\beta
   E'_{\alpha\gam}}\right| 
+  \sum_{\alpha,\gam} \ln \left|1-e^{-\beta
  Z'_{\alpha\gam}}\right| ,
\label{EZ}
\end{multline}
where $E_\gam(\k)$ and $Z_{\gam}(\k)$ [$E'_{\alpha\gam}$ and
  $Z'_{\alpha\gam}$] denote the poles and the zeros of the Green function
  $G(z)$ [$G'(z)$]. By virtue of the definition of the self-energy,
  $Z_{\gam}(\k)$ and $Z'_{\alpha\gam}$ correspond to the  poles of the
 $\tilde\Sigma(z)$ and $\Sigma(z)$, respectively. We show in Appendix
  \ref{app_self} that $\tilde\Sigma(z)$ and $\Sigma(z)$ share the same
  poles, i.e. $\lbrace Z_{\gam}(\k)\rbrace=\lbrace
  Z'_{\alpha\gam}\rbrace$. From Eqs.~(\ref{Omega3},\ref{EZ}), we then
  deduce  
\begin{multline}
\Omega_x(x') = \Omega' - \frac{1}{N} \sum_{\k,\gam} E_\gam(\k)
  \theta[-E_\gam(\k)] \\
+  \frac{1}{N} \sum_{\alpha,\gam} E'_{\alpha\gam}
  \theta[-E'_{\alpha\gam}] 
\label{Omega5}
\end{multline}
($\theta$ is the Heaviside step function) in the zero temperature
limit. Note that the energies $E'_{\alpha\gam}$ are $N_c$ times
degenerate, since $G'$ is the Green function of the whole reference system
($N_c$ clusters). Eq.~(\ref{Omega5}) gives the boson density
\beq
n =  \frac{1}{N} \sum_{\k,\gam} \frac{\partial E_\gam(\k)}{\partial\mu}
\biggr|_{x'} \theta[-E_\gam(\k)] .
\label{density2}
\eeq 
To relate $n$ to the boson density $n_c=-\partial\Omega'/\partial\mu'$ in the
reference system, we use the stationary condition $\partial
\Omega_x(x')/\partial\mu'=0$, 
\bleq
n_c &=& - \frac{1}{N}  \sum_{\k,\gam} \frac{\partial E_\gam(\k)}{\partial\mu'}
   \biggr|_{x} \theta[-E_\gam(\k)] \nonumber \\ &&
+  \frac{1}{N} \sum_{\alpha,\gam} \frac{\partial
  E'_{\alpha\gam}}{\partial \mu'} \biggr|_{x} 
  \theta[-E'_{\alpha\gam}] .
\label{density_c}
\eleq 
This equation can be simplified by noting that the chemical potential $\mu'$
is a mere shift of the excitation energies $E'_{\alpha\gam}$, so that
$E'_{\alpha\gam}+\mu'$ is independent of $\mu'$, 
\beq
\frac{\partial E'_{\alpha\gam}}{\partial \mu'} \biggr|_{x} = -1. 
\label{dEpdmup}
\eeq
Consider now the relation (\ref{Sigma_tilde}) between $G(z)$, $G_0(z)$ and
$\tilde\Sigma(z)$. In this equation, $\tilde\Sigma(z)$ is a function of
$z+\mu'$, while $G_0(z)$ and $G(z)$ are functions of $z+\mu$. It follows that
$E_\gam(\k)+\mu$ is a function of $\mu-\mu'$, which gives
\beq
\frac{\partial E_{\gam}(\k)}{\partial \mu}\biggr|_{x'} + 1 = - \frac{\partial
  E_{\gam}(\k)}{\partial \mu'} \biggr|_{x} .
\label{dEdmu} 
\eeq 
From Eqs.~(\ref{density2}-\ref{dEdmu}), we deduce 
\beq
n = n_c + \frac{1}{N} \sum_{\alpha,\gam} \theta[-E'_{\alpha\gam}]
        - \frac{1}{N} \sum_{\k,\gam}  \theta[-E_\gam(\k)] .
\eeq 
In the Mott phase, the coupling between clusters will transform the discrete
excitation energies $E'_{\alpha\gam}$ into energy bands
$E_\gam(\k)$. Since the gap in the excitation spectrum remains nonzero, the
number of negative energies cannot change, so that $n=n_c$. 

Clearly, and this
is confirmed by our numerical calculations, the important point here is to
take the chemical potential $\mu'$ of the cluster as a variational
parameter. Other parameters, such as the intersite hopping matrix $\hat t'$,
do not have to be considered as variational in order to obtain a consistent
result for the boson density. This point has also been stressed in
Ref.~\onlinecite{Aichhorn05}.

\section{Numerical results in one dimension} 
\label{sec_qunu}

In this section, we present numerical results for 1D systems. 

The simplest reference system consists of single-site clusters ($L=1$). The
only variational parameter is then the chemical potential $x'=\mu'$. A
single-site cluster can be solved analytically exactly. The state with $p\geq
0$ particles is an eigenstate with the energy $\eps_p=-\mu'p+(U/2)p(p-1)$. This
yields the partition function
\beq
e^{-\beta\Omega'} =\sum_{p=0}^\infty e^{-\beta\eps_p} \to e^{-\beta
  \eps_{n_c}} \;\;\; (T\to 0) .  
\eeq 
The number of bosons $n_c$ in the ground state is obtained from
$\eps_{n_c}=\min_p\eps_p$. This condition leads to $n_c-1\leq\mu'/U\leq n_c$
if $\mu'>-U$, and $n_c=0$ if $\mu'<0$. In the following, we consider
only the MI with one boson per site, $n=n_c=1$, which requires
$0<\mu'<U$. The zero-temperature Green function is local in space and
reads\cite{Sengupta05} 
\beq
G'(i\wn) = \frac{-1}{i\wn+\mu'} + \frac{2}{i\wn+\mu'-U} .
\eeq 

The self-energy $\Sigma(i\wn) = i\wn+\mu'-G'{}^{-1}(i\wn)$ is local in real
space and diagonal in reciprocal space: $\tilde\Sigma=\Sigma$. From
Eq.~(\ref{Sigma_tilde}) and $G_0^{-1}(k,i\wn)=i\wn+\mu-\eps_k$, we obtain 
\bleq
G(k,i\wn) &=& \frac{G'(i\wn)}{1-(\Delta\mu'+\eps_k)G'(i\wn)} \nonumber \\
&=& \frac{1-z_k}{i\wn-E^-_k} +  \frac{z_k}{i\wn-E^+_k} ,
\eleq 
where
\begin{multline}
E^\pm_k = -\mu + \half \left(-\Delta\mu'+\eps_k+U \right) \\ 
\pm \half \left[ U^2 + 6U(\Delta\mu'+\eps_k) + (\Delta\mu'+\eps_k)^2
  \right]^{1/2}
\label{single1}  
\end{multline}
and $\Delta\mu'=\mu'-\mu$. Given that $-\mu',E^-_k<0$ and $U-\mu',E^+_k>0$
(see below), Eq.~(\ref{Omega5}) gives 
\beq 
\Omega_x(\mu') = -2\mu'- \frac{1}{N} \sum_k E^-_k . 
\label{single2}
\eeq 

\begin{figure}
\centerline{\includegraphics[width=7cm]{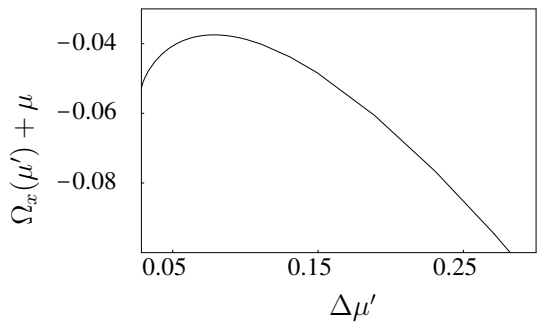}}
\caption{Grand potential $\Omega_x(\mu')$ {\it vs.} $\Delta\mu'=\mu'-\mu$ in
  1D for $t=0.1$ (single-site clusters). In all figures $U=1$. }
\label{fig_omega}
\end{figure} 

\begin{figure}
\centerline{\includegraphics[width=7cm]{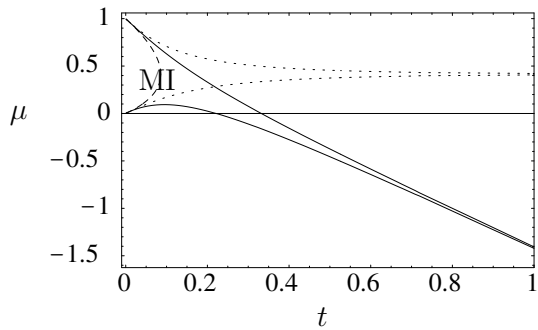}}
\caption{Phase diagram in 1D obtained from a single-site cluster. 
  The solid lines show the
  boundaries between the $n=1$ Mott insulator (MI) and the surrounding
  superfluid phase. The dashed line is the CPT result. The dotted lines show
  the values of $\mu'$ at the Mott-insulator--to--superfluid transition.} 
\label{fig_1D_single} 
\end{figure}

For $\Delta\mu'=0$, we recover the single-site CPT results obtained
earlier.\cite{VanOosten01,Konabe04,Sengupta05} A finite $\Delta\mu'$ does not
change the structure of the Green function $G(k,i\wn)$. The excitation
spectrum reveals the generic characteristics of the MI-SF
transition.\cite{Fisher89} There are two excitation branches,
$E_k^+>0$ and $E_k^-<0$, which coincide with the cluster excitation energies
$-\mu'$ and $U-\mu'$ in the limit $t\to 0$. The dispersion of
$E_k^\pm$ increases with $t$, which leads to a decrease of the Mott gap
$E^+_{k=0}-E^-_{k=0}$.  (See, for instance, the figures in
Ref.~\onlinecite{Sengupta05} for the case $\Delta\mu'=0$.) By varying the
chemical potential $\mu$ (with $t$ and 
$U$ fixed), one induces a transition from the commensurate incompressible MI to
the incommensurate compressible SF when $E^+_{k=0}$ or $E^-_{k=0}$ vanishes.
This transition is density driven and its critical behavior mean-field
like. At the tip of the Mott lobe shown in
Fig.~\ref{fig_1D_single}, $E^+_{k=0}$ and $E^-_{k=0}$ vanish simultaneously,
i.e. the Mott gap closes. This transition occurs at fixed density; it is
driven by phase fluctuations and is in the universality class of the $(d+1)$ XY
model (with $d$ the space dimension).\cite{Fisher89}

Within the VCPT, one has to determine $\Delta\mu'$ from the
stationary condition $\partial\Omega_x(\mu')/\partial\mu'=0$, 
\beq
\frac{1}{N}\sum_k \frac{3U+\Delta\mu'+\eps_k}{ \left[ U^2
  +6U(\Delta\mu'+\eps_k) + (\Delta\mu'+\eps_k)^2 \right]^{1/2}} = 3. 
\label{single3}
\eeq 
Using Eq.~(\ref{single3}), one can verify that the boson density
$n=-\partial\Omega/\partial\mu$ obtained from Eq.~(\ref{single2}) equals
that of the cluster, $n=n_c=1$, in agreement with the general
proof given in Sec.~\ref{subsec_mup}. In order for the square root in
Eqs.~(\ref{single1},\ref{single3}) to be defined, $\Delta\mu'$ should satisfy
$\Delta\mu'\leq D-(3+2\sqrt{2})U$ or $\Delta\mu'\geq D-(3-2\sqrt{2})U$, where
$D=-\eps_{k=0}=zt$. (Note that the Mott gap vanishes when $\Delta\mu'= D-(3\pm
2\sqrt{2})U$.) The grand potential $\Omega_x(\mu')$ is shown in
Fig.~\ref{fig_omega} for $t/U=0.1$. We see that the stationary point of
$\Omega_x(\mu')$ is a maximum with respect to variations of $\mu'$. This
turns out to be always true regardless of the space dimension or the number of
sites in the cluster. Since we expect $\Omega_x(x')$ to be a minimum with
respect to other variational parameters -- for instance the intracluster
hopping amplitude $t'$ -- the 
stationary point of the grand potential $\Omega_x(x')$ will be in general a
saddle point for larger clusters ($L\geq 2$). 

Once $\Delta\mu'$ is known, the
values of the chemical potential $\mu$ for which the Mott state becomes
unstable are obtained from $E_{k=0}^+=0$ and $E_{k=0}^-=0$. 
The lowest value of $\Delta\mu'$ shown in Fig.~\ref{fig_omega} 
corresponds to $\Delta\mu' = D-(3-2\sqrt{2})U$. As $t/U$ increases, this value
moves closer to the stationary point, and coincides with it when the Mott gap
closes. This signals a transition to the superfluid state with a density
$n=1$. For larger values of $t$, the ground state is superfluid for any value
of $\mu$ (provided the boson density remains finite). 

The phase diagram of the 1D Bose-Hubbard model in Fig.~\ref{fig_1D_single}
shows that the results obtained from the CPT and the VCPT differ
drastically even for single-site clusters. Whereas the CPT gives the
round-shape Mott lobe  characteristic of 
mean-field theories -- and independent of the dimension but for 
a trivial dependence on the number $z$ of nearest neighbors --, the VCPT
qualitatively reproduces the shape of the Mott lobe in
1D.\cite{Kuhner98,Kuhner00,Elstner99} In particular, it yields a reentrant
behavior of the MI state as $t$ is increased with $\mu\sim 0.05 U$
fixed, and a very pointed shape around the lobe tip. For
sufficiently large values of $t$, $\mu'$ comes near $U/2$, while the decrease
of $\mu$ appears to be tied to the bottom of the free boson dispersion
$\eps_k$, i.e. $\mu\sim -2t+\const$. The gap closes very slowly as $t$
increases, and the MI disappears for $t/U\sim 2$ (not shown in
Fig.~\ref{fig_1D_single}).

Note that the pointed shape of the Mott lobe in 1D is usually
attributed to the slow decrease of the Mott gap, 
\beq
E_{k=0}^+ - E_{k=0}^- \sim \exp \left( - \frac{\const}{\sqrt{t_c-t}} \right),
\eeq 
near the Berezinskii-Kosterlitz-Thouless (BKT) transition taking place at the
lobe tip ($t=t_c$).\cite{Kuhner00,Elstner99} The fact that the single-site
VCPT reproduces the correct overall shape of the Mott tip, while the physics
of the BKT transition is clearly out of reach of this method, is quite
remarkable. 

For larger clusters ($L\geq 2$), the ground state and the grand potential
$\Omega'$, as well as the single-particle Green function $G'$, can be obtained
numerically using the Lanczos method.\cite{Senechal02} The intracluster
hopping amplitude $t'=t$ is held fixed, and only the chemical
potential $\mu'$ is taken as a variational parameter. As in the case $L=1$,
the stationary point $\partial\Omega_x(\mu')/\partial\mu'=0$ is a
maximum. For a given $\mu$, not all values of $\mu'$ are physically
acceptable. On the one hand, $\mu'$ has to be such that $n_c=1$ (we restrict
ourselves to the $n=1$ MI). On the other hand, we have to ensure
that the ground state is stable. For $L=1$, we saw that for
certain values of $\mu'$, the single-particle excitation energies are
complex. This instability also shows up as a disagreement between the boson
density $n$ obtained from the trace of $G$ [Eq.~(\ref{density3})] and
$n_c$. We use this latter criterion to verify the stability of the ground
state, since it does not require to obtain the
excitation energies while minimizing the grand potential with respect to
$\mu'$. 

\begin{figure}
\centerline{\includegraphics[width=7cm]{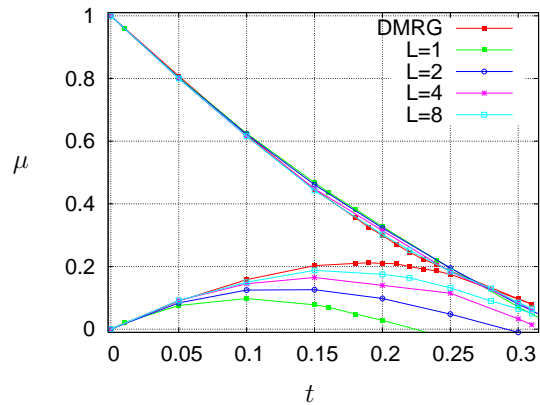}}
\caption{(color online) Phase diagram in 1D obtained from a $L$-site cluster. 
  The (red) squares show the results from the density-matrix renormalization
  group.\cite{Kuhner00} }
\label{fig_dia_1D} 
\end{figure}

The phase diagram of the 1D Bose-Hubbard model is shown in
Fig.~\ref{fig_dia_1D} together with the results obtained from the
density-matrix renormalization group (DMRG).\cite{Kuhner00} Single-site
clusters give a good approximation to the upper boundary of the $n=1$ MI. By
contrast, the lower boundary obtained within the VCPT strongly depends on the
number of sites in the cluster. For $L\gtrsim 8$, the VCPT reproduces fairly
accurately the results obtained from the DMRG. 

\begin{figure}[t]
\centerline{\includegraphics[width=5cm]{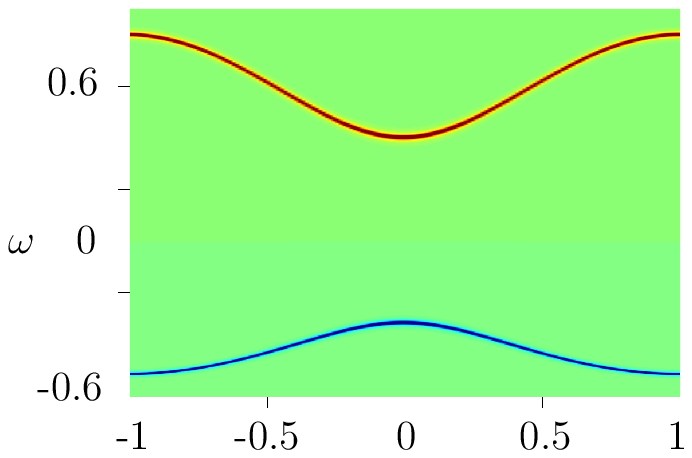}}
\vspace{0.5cm} 
\centerline{\includegraphics[width=5cm]{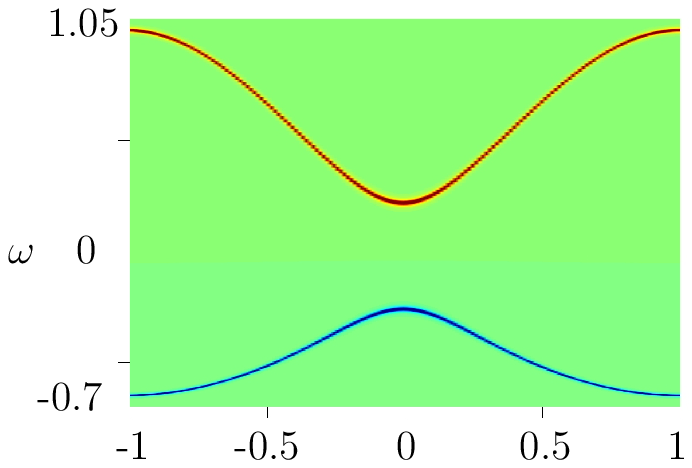}}
\vspace{0.5cm} 
\centerline{\includegraphics[width=5cm]{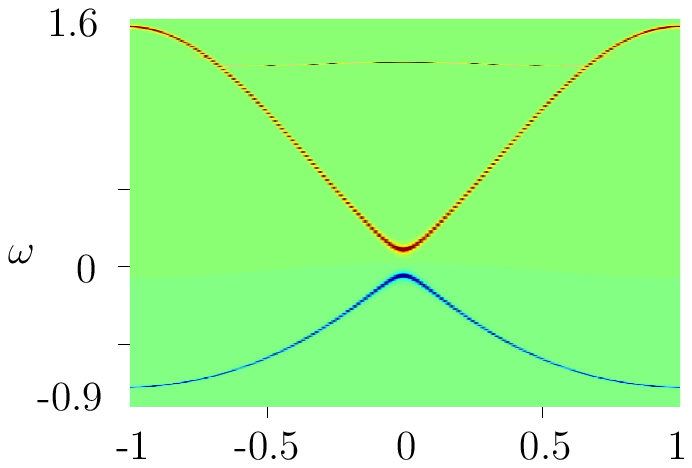}}
\vspace{0.5cm} 
\centerline{\includegraphics[width=5cm]{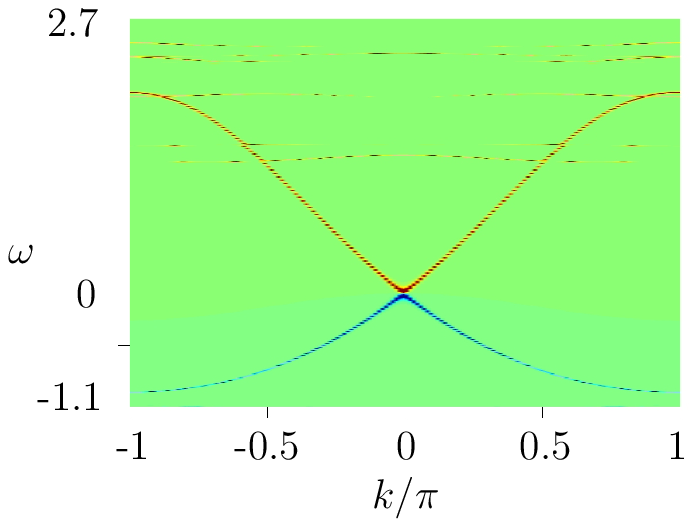}}
\caption{(color online) Spectral function $A(k,\w)=-\pi^{-1} {\rm Im}
  G(k,\w+i\eta)$ obtained from 4-site clusters: $t/U=0.05$, 0.1,
  0.2 and 0.3 (from top to bottom).} 
\label{fig_spectre}
\end{figure}

We show in Fig.~\ref{fig_spectre} the spectral function $A(k,\w)=-\pi^{-1}
{\rm Im} G(k,\w+i\eta)$ ($\w$ real). The spectrum consists of two well defined
dispersion branches separated by the Mott gap. We see clearly how the Mott gap
closes as we move closer to the Mott lobe tip. Note that besides the two main
excitation energies, the spectral function exhibits additional (weaker)
structures at positive energies when the Mott gap is small (see the two bottom
graphs in Fig.~\ref{fig_spectre}); we do not know whether these have a true
physical meaning or are due to the finite accuracy of our numerical
calculations. 

\begin{figure}
\centerline{\includegraphics[bb=60 322 280 507,width=6cm]{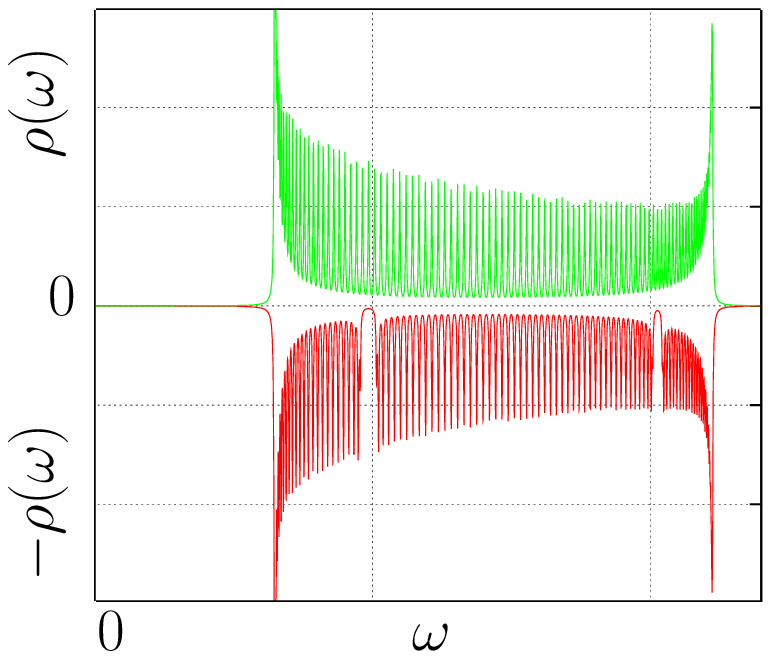}}
\caption{(color online) Comparison between the single-particle density of
  states ($\omega > 0$) obtained from the translation invariant self-energy
  $\tilde\Sigma(k,z)$ (upper (green) curve $\rho(\omega)$) with that obtained
  from the self-energy $\Sigma(k,k',z)$ (lower (red) curve
  $-\rho(\omega)$). The plots are obtained from 4-site clusters. }
\label{fig_gap} 
\end{figure} 

As discussed in Sec.~\ref{sec_vcpt}, by making use of the translation
invariant self-energy $\tilde\Sigma$, we obtain a boson dispersion free of
spurious gaps coming from the periodicity of the reference system
(Fig.~\ref{fig_spectre}). This is illustrated in Fig.~\ref{fig_gap}, where we
compare the single-particle density of states 
$\rho(\w)=\int \frac{dk}{2\pi} A(k,\omega)$ with that obtained from the
non-translation-invariant self-energy $\Sigma(k,k',z)$. In the latter case,
two gaps arising from the periodicity of the 4-site clusters are clearly
visible.

\section{Conclusion}

The VCPT,\cite{Potthoff03a,Potthoff03b,Potthoff03c,Dahnken04}
which was previously applied to strongly-correlated fermion
systems,\cite{Dahnken04,Aichhorn04,Senechal05,Aichhorn05,Danhken05} can be
extended to boson systems. We propose a modification of the original
formulation which ensures that the final results are translation invariant
despite the fact that the reference system breaks translation invariance. This
translation invariant VCPT is not restricted to boson systems but should also
apply to fermion systems.  

The results obtained for the 1D Bose-Hubbard model indicate that the VCPT is
an efficient method for studying the Mott-insulator--to--superfluid transition
in boson systems. We stress the importance of taking the chemical
potential $\mu'$ of the cluster as a variational parameter. This ensures
that the approach is thermodynamically consistent
($n=-\partial\Omega/\partial\mu= 
-\Tr(G)/(N\beta)$), and that the boson density $n$ is the same in the system
and the reference system. The grand
potential is found to be a maximum with respect to variation of $\mu'$,
which implies that in general it will correspond to a saddle point when
several variational parameters are considered.  
Even for one-site clusters, the VCPT and CPT --
where $\mu'=\mu$ is not a variational parameter -- differ drastically. Whereas
the CPT gives the usual dimension-independent mean-field results, the VCPT
reproduces the characteristic pointed shape and reentrant behavior of the Mott
lobe in 1D. 

The extension of our results to higher dimension does not raise any difficulty
and will be discussed elsewhere. More subtle is the application of the VCPT to
the superfluid phase. In order to describe a broken gauge-symmetry phase
($\mean{\psi_\r}\neq 0$), one should add to the reference system Hamiltonian a
field $h'_\r$ that couples to the boson operator $\psi_\r$. The set $x'$ of
variational parameters will therefore at least include the chemical potential
$\mu'$ and the field $h'$. A finite value of $h'$ at the stationary point
implies a nonzero condensate $\mean{\psi_\r}$ and superfluidity. Whether or
not the VCPT will satisfy the Hugenholz-Pines theorem and thus correctly
describe the gapless Bogoliubov sound mode should determine the applicability
of the VCPT to the superfluid phase.

\begin{acknowledgments} 
W.~K. thanks the EPSRC for support through the Grant GR/S18571/01.          
\end{acknowledgments} 

\appendix

\section{Derivation of Eq.~(\ref{subadd})}
\label{app_subadd} 

\begin{figure}
\centerline{\includegraphics[width=4cm]{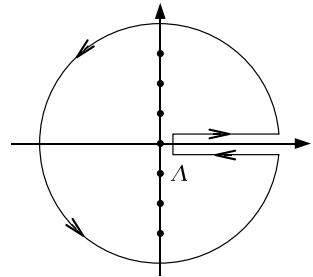}}
\caption{Integration contour used in Eq.~(\ref{app11}). The black dots denote
  the bosonic Matsubara frequencies $i\wn$. }
\label{fig_contour1}
\end{figure}

Let us consider the integral 
\bleq
I &=& \oint \frac{dz}{2i\pi} n_B(z) \tr_c \ln\left(1- \frac{\Delta
  h_0(\K)}{z-y} \right) e^{z \eta} \nonumber \\ 
&=& \oint \frac{dz}{2i\pi} n_B(z) \sum_{\alpha=1}^L \ln\left(1-
\frac{\lambda_\alpha(\K)}{z-y} \right) e^{z \eta} ,
\label{app11}
\eleq
where $\lambda_\alpha(\K)$ ($\alpha\in [1,L]$) are the eigenvalues of $\Delta
h_0(\K)$. $n_B(z)=(e^{\beta z}-1)^{-1}$ is the Bose-Einstein distribution
function. If we choose $y>\max_{\alpha,\K} |\lambda_\alpha(\K)|$, the branch
cut of the logarithm in Eq.~(\ref{app11}) does not extend to the origin $z=0$,
and we can evaluate $I$ using the contour shown in
Fig.~\ref{fig_contour1} where $0<\Lambda\leq
y-\max_{\alpha,\K}|\lambda_\alpha(\K)|$. From the residue theorem, we then
obtain 
\bleq
I &=& \frac{1}{\beta} \sum_{\alpha,\wn}  \ln\left(1-
\frac{\lambda_\alpha(\K)}{i\wn-y} \right) e^{i\wn \eta} \nonumber \\ 
&=& \int_\Lambda^\infty \frac{d\eps}{2i\pi} n_B(\eps) \sum_{\alpha=1}^L \left[ 
\ln\left(1 - \frac{\lambda_\alpha(\K)}{\eps+i\eta-y} \right) - {\rm c.c.}
\right] . \nonumber \\ && 
\label{app12}
\eleq
The factor $n_B(z)e^{z\eta}$ ensures that the part of the contour at infinity
does not contribute to the integral. An integration by part gives
\bleq
I &=&  - \int_\Lambda^\infty \frac{d\eps}{2i\pi} \frac{1}{\beta}
\ln\left(1-e^{-\beta\eps}\right) \nonumber \\ && \times 
\sum_{\alpha=1}^L \frac{\partial}{\partial \eps} \left[ 
\ln\left(1 - \frac{\lambda_\alpha(\K)}{\eps+i\eta -y} \right) - {\rm c.c.}
\right] \nonumber \\ 
&=& \frac{1}{\beta} \sum_{\alpha=1}^L
\ln\left(1-e^{-\beta(y+\lambda_\alpha(\K))}\right) 
- \frac{L}{\beta} \ln\left(1-e^{-\beta y}\right) . \nonumber \\ && 
\label{app13}
\eleq
Eq.~(\ref{subadd}) follows from (\ref{app12}) and (\ref{app13}). 

\section{Derivation of Eqs.~(\ref{EZ})}
\label{app_int} 

\begin{figure}[t]
\centerline{\includegraphics[width=4cm]{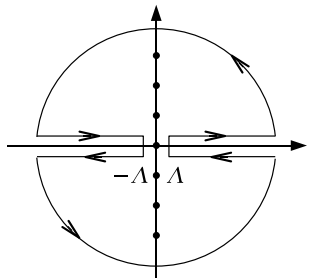}}
\caption{Integration contour used in Eq.~(\ref{app21}). }
\label{fig_contour2}
\end{figure}

Let us consider the integral 
\bleq
I &=& \oint \frac{dz}{2i\pi} n_B(z) \tr\ln[-G'(z)] e^{z\eta} \nonumber
\\ &=&\oint \frac{dz}{2i\pi} n_B(z) \sum_{\alpha=1}^L
\ln[-G'_\alpha(z)] e^{z\eta} ,
\label{app21}
\eleq
where $G'_\alpha(z)$ ($\alpha\in [1,L]$) are the eigenvalues of $G'(z)$. The
(exact) bosonic Green function can be written as a sum of simple poles,
\beq
G'_\alpha(z) = \sum_\gam \frac{r_{\alpha\gam}}{z-E'_{\alpha\gam}} ,
\label{app22}
\eeq
where $\sgn(r_{\alpha\gam})=\sgn(E'_{\alpha\gam})$. The excitation energies
$E'_{\alpha\gam}$ are nonzero in the Mott phase. As a result
\beq
-G'_\alpha(z=0) = \sum_\gam \frac{r_{\alpha\gam}}{E'_{\alpha\gam}} > 0 
\eeq
and the branch cut of the logarithm in Eq.~(\ref{app21}) does not extend to
the origin. We can therefore evaluate $I$ using the contour shown in
Fig.~\ref{fig_contour2},
\bleq
I &=& \frac{1}{\beta} \sum_{\alpha,\wn} \ln[-G'_\alpha(i\wn)] e^{i\wn \eta} 
\nonumber \\ 
&=& \left( \int_\Lambda^\infty + \int_{-\infty}^{-\Lambda} \right)
\frac{d\eps}{2i\pi} n_B(\eps) \nonumber \\ && \times \sum_\alpha \left\lbrace
\ln[-G'_\alpha(\eps+i\eta)] - {\rm c.c.} \right\rbrace \nonumber \\ 
&=&  - \left( \int_\Lambda^\infty + \int_{-\infty}^{-\Lambda} \right)
\frac{d\eps}{2i\pi} \frac{1}{\beta} \ln\left|1-e^{-\beta\eps}\right| \nonumber
     \\ && \times \sum_\alpha 
  [ G'_\alpha{}^{-1}(\eps+i\eta) \partial_\eps G'_\alpha(\eps+i\eta) - {\rm
    c.c.}] ,
\label{app23}
\eleq
where $0<\Lambda<\min_{\alpha,\gam}(|E'_{\alpha\gam}|,|Z'_{\alpha\gam}|)$, with
$Z'_{\alpha\gam}$ the zeros of $G'(z)$. The last line in
(\ref{app23}) is obtained by integrating by part. The factor inside the 
last brackets vanishes unless $\eps$ is near a pole or a zero of
$G'_\alpha(\eps)$. Near a pole, $G'_\alpha(\eps)\simeq
r_{\alpha\gam}/(\eps+i\eta-E'_{\alpha\gam})$, and
$[\cdots]=2i\pi\delta(\eps-E'_{\alpha\gam})$. The zeros of
$G'_\alpha(z)$ are given by the poles of the self-energy $\Sigma(z)$, which
shares the same analytical properties as $G'(z)$ and can therefore be written
as a sum of simple poles as in Eq.~(\ref{app22}) (see Appendix \ref{app_self}
for a more detailed discussion). It follows that $G'_\alpha(z)$ has only simple
zeros $Z'_{\alpha\gam}$ and can be approximated by $G'_\alpha(z)\simeq
s_{\alpha\gam}(z-Z'_{\alpha\gam})$ near $z=Z'_{\alpha\gam}$. The factor inside
the brackets in Eq.~(\ref{app23}) then gives
$-2i\pi\delta(\eps-Z'_{\alpha\gam})$. We deduce 
\beq
I = - \frac{1}{\beta} \sum_{\alpha,\gam} \ln\left|1-e^{-\beta
  E'_{\alpha\gam}} \right|  + \frac{1}{\beta} \sum_{\alpha,\gam}
\ln\left|1-e^{-\beta Z'_{\alpha\gam}}\right| .
\label{app24}
\eeq 
The second of Eqs.~(\ref{EZ}) follows from (\ref{app23}) and (\ref{app24}). The
first one can be derived similarly. Analog results for fermionic systems can
be found in Ref.~\onlinecite{Potthoff03b}.

\section{Cluster self-energy $\Sigma(z)$} 
\label{app_self}

In this appendix, we discuss two important properties of the cluster
self-energy $\Sigma(z)$. 

\subsection{$\lim_{|z|\to\infty}\Sigma(z)$}

The cluster self-energy $\Sigma(z)$ tends to a constant value equal the
Hartree-Fock value $2Un$ when $|z|\to\infty$. A similar result holds for the
fermionic Hubbard model: see for instance Appendix A in
Ref.~\onlinecite{Vilk97}; the derivation given in this paper can be
straightforwardly extended to the Bose-Hubbard model. 

\subsection{Poles of $\Sigma(z)$ and $\tilde\Sigma(z)$} 

The (matrix) self-energy $\Sigma(z)$ shares the same analytical properties as
the Green function $G'(z)$. Its Lehmann representation 
\beq
\Sigma(z) = \Sigma(\infty) + \sum_{\alpha,\gam} \ket{\alpha,\gam}
\frac{q_{\alpha\gam}}{z-Z_{\alpha\gam}}\bra{\alpha,\gam} 
\eeq
(we use a bra-ket notation) leads to 
\bleq
\tilde\Sigma(\k,z) &=& \braket{\k|\Sigma(z)}{\k} \nonumber \\ 
&=& \Sigma(\infty) + \sum_{\alpha,\gam}
|\braket{\k}{\alpha,\gam}|^2 
\frac{q_{\alpha\gam}}{z-Z_{\alpha\gam}}  ,
\eleq 
where $\Sigma(\infty)=2Un$. Since both $\lbrace\ket{\k}\rbrace$ and
$\lbrace\ket{\alpha,\gam}\rbrace$ span the whole Hilbert space,
$\braket{\k}{\alpha,\gam}$ cannot vanish for all $\k$. This implies that all
poles $Z_{\alpha\gam}$ of $\Sigma(z)$ also show up as poles of
$\tilde\Sigma(z)$.  


\begin{thebibliography}{43}
\expandafter\ifx\csname natexlab\endcsname\relax\def\natexlab#1{#1}\fi
\expandafter\ifx\csname bibnamefont\endcsname\relax
  \def\bibnamefont#1{#1}\fi
\expandafter\ifx\csname bibfnamefont\endcsname\relax
  \def\bibfnamefont#1{#1}\fi
\expandafter\ifx\csname citenamefont\endcsname\relax
  \def\citenamefont#1{#1}\fi
\expandafter\ifx\csname url\endcsname\relax
  \def\url#1{\texttt{#1}}\fi
\expandafter\ifx\csname urlprefix\endcsname\relax\def\urlprefix{URL }\fi
\providecommand{\bibinfo}[2]{#2}
\providecommand{\eprint}[2][]{\url{#2}}

\bibitem[{\citenamefont{Fisher et~al.}(1989)\citenamefont{Fisher, Weichman,
  Grinstein, and Fisher}}]{Fisher89}
\bibinfo{author}{\bibfnamefont{M.~P.~A.} \bibnamefont{Fisher}},
  \bibinfo{author}{\bibfnamefont{P.~B.} \bibnamefont{Weichman}},
  \bibinfo{author}{\bibfnamefont{G.}~\bibnamefont{Grinstein}},
  \bibnamefont{and} \bibinfo{author}{\bibfnamefont{D.~S.}
  \bibnamefont{Fisher}}, \bibinfo{journal}{Phys. Rev. B}
  \textbf{\bibinfo{volume}{40}}, \bibinfo{pages}{546} (\bibinfo{year}{1989}).

\bibitem[{\citenamefont{Jaksch et~al.}(1998)\citenamefont{Jaksch, Bruder,
  Cirac, Gardiner, and Zoller}}]{Jaksch98}
\bibinfo{author}{\bibfnamefont{D.}~\bibnamefont{Jaksch}},
  \bibinfo{author}{\bibfnamefont{C.}~\bibnamefont{Bruder}},
  \bibinfo{author}{\bibfnamefont{J.~I.} \bibnamefont{Cirac}},
  \bibinfo{author}{\bibfnamefont{C.~W.} \bibnamefont{Gardiner}},
  \bibnamefont{and} \bibinfo{author}{\bibfnamefont{P.}~\bibnamefont{Zoller}},
  \bibinfo{journal}{Phys. Rev. Lett.} \textbf{\bibinfo{volume}{81}},
  \bibinfo{pages}{3108} (\bibinfo{year}{1998}).

\bibitem[{\citenamefont{Greiner et~al.}(2002)\citenamefont{Greiner, Mandel,
  Esslinger, H\"ansch, and Bloch}}]{Greiner02}
\bibinfo{author}{\bibfnamefont{M.}~\bibnamefont{Greiner}},
  \bibinfo{author}{\bibfnamefont{O.}~\bibnamefont{Mandel}},
  \bibinfo{author}{\bibfnamefont{T.}~\bibnamefont{Esslinger}},
  \bibinfo{author}{\bibfnamefont{T.~W.} \bibnamefont{H\"ansch}},
  \bibnamefont{and} \bibinfo{author}{\bibfnamefont{I.}~\bibnamefont{Bloch}},
  \bibinfo{journal}{Nature} \textbf{\bibinfo{volume}{415}}, \bibinfo{pages}{39}
  (\bibinfo{year}{2002}).

\bibitem[{\citenamefont{St\"oferle et~al.}(2004)\citenamefont{St\"oferle,
  Moritz, Schori, K\"ohl, and Esslinger}}]{Stoferle04}
\bibinfo{author}{\bibfnamefont{T.}~\bibnamefont{St\"oferle}},
  \bibinfo{author}{\bibfnamefont{H.}~\bibnamefont{Moritz}},
  \bibinfo{author}{\bibfnamefont{C.}~\bibnamefont{Schori}},
  \bibinfo{author}{\bibfnamefont{M.}~\bibnamefont{K\"ohl}}, \bibnamefont{and}
  \bibinfo{author}{\bibfnamefont{T.}~\bibnamefont{Esslinger}},
  \bibinfo{journal}{Phys. Rev. Lett.} \textbf{\bibinfo{volume}{92}},
  \bibinfo{pages}{130403} (\bibinfo{year}{2004}).

\bibitem[{\citenamefont{Gerbier et~al.}(2005)\citenamefont{Gerbier, Widera,
  F\"olling, Mandel, Gericke, and Bloch}}]{Gerbier05}
\bibinfo{author}{\bibfnamefont{F.}~\bibnamefont{Gerbier}},
  \bibinfo{author}{\bibfnamefont{A.}~\bibnamefont{Widera}},
  \bibinfo{author}{\bibfnamefont{S.}~\bibnamefont{F\"olling}},
  \bibinfo{author}{\bibfnamefont{O.}~\bibnamefont{Mandel}},
  \bibinfo{author}{\bibfnamefont{T.}~\bibnamefont{Gericke}}, \bibnamefont{and}
  \bibinfo{author}{\bibfnamefont{I.}~\bibnamefont{Bloch}},
  \bibinfo{journal}{Phys. Rev. Lett.} \textbf{\bibinfo{volume}{95}},
  \bibinfo{pages}{050404} (\bibinfo{year}{2005}).

\bibitem[{\citenamefont{Rokhsar and Kotliar}(1991)}]{Rokhsar91}
\bibinfo{author}{\bibfnamefont{D.~S.} \bibnamefont{Rokhsar}} \bibnamefont{and}
  \bibinfo{author}{\bibfnamefont{B.~G.} \bibnamefont{Kotliar}},
  \bibinfo{journal}{Phys. Rev. B} \textbf{\bibinfo{volume}{44}},
  \bibinfo{pages}{10328} (\bibinfo{year}{1991}).

\bibitem[{\citenamefont{Krauth et~al.}(1992)\citenamefont{Krauth, Caffarel, and
  Bouchaud}}]{Krauth92}
\bibinfo{author}{\bibfnamefont{W.}~\bibnamefont{Krauth}},
  \bibinfo{author}{\bibfnamefont{M.}~\bibnamefont{Caffarel}}, \bibnamefont{and}
  \bibinfo{author}{\bibfnamefont{J.-P.} \bibnamefont{Bouchaud}},
  \bibinfo{journal}{Phys. Rev. B} \textbf{\bibinfo{volume}{45}},
  \bibinfo{pages}{3137} (\bibinfo{year}{1992}).

\bibitem[{\citenamefont{Schroll et~al.}(2004)\citenamefont{Schroll, Manquardt,
  and Bruder}}]{Schroll04}
\bibinfo{author}{\bibfnamefont{C.}~\bibnamefont{Schroll}},
  \bibinfo{author}{\bibfnamefont{F.}~\bibnamefont{Manquardt}},
  \bibnamefont{and} \bibinfo{author}{\bibfnamefont{C.}~\bibnamefont{Bruder}},
  \bibinfo{journal}{Phys. Rev. A} \textbf{\bibinfo{volume}{70}},
  \bibinfo{pages}{053609} (\bibinfo{year}{2004}).

\bibitem[{\citenamefont{K\"uhner and Monien}(1998)}]{Kuhner98}
\bibinfo{author}{\bibfnamefont{T.~D.} \bibnamefont{K\"uhner}} \bibnamefont{and}
  \bibinfo{author}{\bibfnamefont{H.}~\bibnamefont{Monien}},
  \bibinfo{journal}{Phys. Rev. B} \textbf{\bibinfo{volume}{58}},
  \bibinfo{pages}{R14741} (\bibinfo{year}{1998}).

\bibitem[{\citenamefont{K\"uhner et~al.}(2000)\citenamefont{K\"uhner, White,
  and Monien}}]{Kuhner00}
\bibinfo{author}{\bibfnamefont{T.~D.} \bibnamefont{K\"uhner}},
  \bibinfo{author}{\bibfnamefont{S.~R.} \bibnamefont{White}}, \bibnamefont{and}
  \bibinfo{author}{\bibfnamefont{H.}~\bibnamefont{Monien}},
  \bibinfo{journal}{Phys. Rev. B} \textbf{\bibinfo{volume}{61}},
  \bibinfo{pages}{12474} (\bibinfo{year}{2000}).

\bibitem[{\citenamefont{Roth and Burnett}(2003{\natexlab{a}})}]{Roth03a}
\bibinfo{author}{\bibfnamefont{R.}~\bibnamefont{Roth}} \bibnamefont{and}
  \bibinfo{author}{\bibfnamefont{K.}~\bibnamefont{Burnett}},
  \bibinfo{journal}{Phys. Rev. A} \textbf{\bibinfo{volume}{67}},
  \bibinfo{pages}{031602(R)} (\bibinfo{year}{2003}{\natexlab{a}}).

\bibitem[{\citenamefont{Roth and Burnett}(2003{\natexlab{b}})}]{Roth03b}
\bibinfo{author}{\bibfnamefont{R.}~\bibnamefont{Roth}} \bibnamefont{and}
  \bibinfo{author}{\bibfnamefont{K.}~\bibnamefont{Burnett}},
  \bibinfo{journal}{Phys. Rev. A} \textbf{\bibinfo{volume}{68}},
  \bibinfo{pages}{023604} (\bibinfo{year}{2003}{\natexlab{b}}).

\bibitem[{\citenamefont{Batrouni et~al.}(1992)\citenamefont{Batrouni,
  Scalettar, and Zimanyi}}]{Batrouni90}
\bibinfo{author}{\bibfnamefont{G.~G.} \bibnamefont{Batrouni}},
  \bibinfo{author}{\bibfnamefont{R.~T.} \bibnamefont{Scalettar}},
  \bibnamefont{and} \bibinfo{author}{\bibfnamefont{G.~T.}
  \bibnamefont{Zimanyi}}, \bibinfo{journal}{Phys. Rev. Lett.}
  \textbf{\bibinfo{volume}{65}}, \bibinfo{pages}{1765} (\bibinfo{year}{1992}).

\bibitem[{\citenamefont{Batrouni and Scalettar}(1992)}]{Batrouni92}
\bibinfo{author}{\bibfnamefont{G.~G.} \bibnamefont{Batrouni}} \bibnamefont{and}
  \bibinfo{author}{\bibfnamefont{R.~T.} \bibnamefont{Scalettar}},
  \bibinfo{journal}{Phys. Rev. B} \textbf{\bibinfo{volume}{46}},
  \bibinfo{pages}{9051} (\bibinfo{year}{1992}).

\bibitem[{\citenamefont{Krauth and Trivedi}(1991)}]{Krauth91}
\bibinfo{author}{\bibfnamefont{W.}~\bibnamefont{Krauth}} \bibnamefont{and}
  \bibinfo{author}{\bibfnamefont{N.}~\bibnamefont{Trivedi}},
  \bibinfo{journal}{Europhys. Lett.} \textbf{\bibinfo{volume}{14}},
  \bibinfo{pages}{627} (\bibinfo{year}{1991}).

\bibitem[{\citenamefont{Wessel et~al.}(2004)\citenamefont{Wessel, Alet, Troyer,
  and Batrouni}}]{Wessel04}
\bibinfo{author}{\bibfnamefont{S.}~\bibnamefont{Wessel}},
  \bibinfo{author}{\bibfnamefont{F.}~\bibnamefont{Alet}},
  \bibinfo{author}{\bibfnamefont{M.}~\bibnamefont{Troyer}}, \bibnamefont{and}
  \bibinfo{author}{\bibfnamefont{G.~G.} \bibnamefont{Batrouni}},
  \bibinfo{journal}{Phys. Rev. A} \textbf{\bibinfo{volume}{70}},
  \bibinfo{pages}{053615} (\bibinfo{year}{2004}).

\bibitem[{\citenamefont{Sheshadri et~al.}(1993)\citenamefont{Sheshadri,
  Krishnamurthy, Pandit, and Ramakrishnan}}]{Sheshadri93}
\bibinfo{author}{\bibfnamefont{K.}~\bibnamefont{Sheshadri}},
  \bibinfo{author}{\bibfnamefont{H.~R.} \bibnamefont{Krishnamurthy}},
  \bibinfo{author}{\bibfnamefont{R.}~\bibnamefont{Pandit}}, \bibnamefont{and}
  \bibinfo{author}{\bibfnamefont{T.~V.} \bibnamefont{Ramakrishnan}},
  \bibinfo{journal}{Europhys. Lett.} \textbf{\bibinfo{volume}{22}},
  \bibinfo{pages}{257} (\bibinfo{year}{1993}).

\bibitem[{\citenamefont{Kampf and Zimanyi}(1993)}]{Kampf93}
\bibinfo{author}{\bibfnamefont{A.~P.} \bibnamefont{Kampf}} \bibnamefont{and}
  \bibinfo{author}{\bibfnamefont{G.~T.} \bibnamefont{Zimanyi}},
  \bibinfo{journal}{Phys. Rev. B} \textbf{\bibinfo{volume}{47}},
  \bibinfo{pages}{279} (\bibinfo{year}{1993}).

\bibitem[{\citenamefont{{van Oosten} et~al.}(2001)\citenamefont{{van Oosten},
  {van der Straten}, and Stoof}}]{VanOosten01}
\bibinfo{author}{\bibfnamefont{D.}~\bibnamefont{{van Oosten}}},
  \bibinfo{author}{\bibfnamefont{P.}~\bibnamefont{{van der Straten}}},
  \bibnamefont{and} \bibinfo{author}{\bibfnamefont{H.~T.~C.}
  \bibnamefont{Stoof}}, \bibinfo{journal}{Phys. Rev. A}
  \textbf{\bibinfo{volume}{63}}, \bibinfo{pages}{053601}
  (\bibinfo{year}{2001}).

\bibitem[{\citenamefont{Sachdev}(1999)}]{Sachdev99}
\bibinfo{author}{\bibfnamefont{S.}~\bibnamefont{Sachdev}},
  \emph{\bibinfo{title}{Quantum Phase Transitions}}
  (\bibinfo{publisher}{Cambridge University}, \bibinfo{address}{Cambridge,
  England}, \bibinfo{year}{1999}).

\bibitem[{\citenamefont{Dickerscheid et~al.}(2003)\citenamefont{Dickerscheid,
  {van Oosten}, Denteneer, and Stoof}}]{Dickerscheid03}
\bibinfo{author}{\bibfnamefont{D.~B.~M.} \bibnamefont{Dickerscheid}},
  \bibinfo{author}{\bibfnamefont{D.}~\bibnamefont{{van Oosten}}},
  \bibinfo{author}{\bibfnamefont{P.~J.~H.} \bibnamefont{Denteneer}},
  \bibnamefont{and} \bibinfo{author}{\bibfnamefont{H.~T.~C.}
  \bibnamefont{Stoof}}, \bibinfo{journal}{Phys. Rev. A}
  \textbf{\bibinfo{volume}{68}}, \bibinfo{pages}{043623}
  (\bibinfo{year}{2003}).

\bibitem{Konabe04} S. Konabe, T. Nikuni, and M. Nakamura, 
Phys. Rev. A {\bf 73}, 033621 (2006).

\bibitem[{\citenamefont{Sengupta and Dupuis}(2005)}]{Sengupta05}
\bibinfo{author}{\bibfnamefont{K.}~\bibnamefont{Sengupta}} \bibnamefont{and}
  \bibinfo{author}{\bibfnamefont{N.}~\bibnamefont{Dupuis}},
  \bibinfo{journal}{Phys. Rev. A} \textbf{\bibinfo{volume}{71}},
  \bibinfo{pages}{033629} (\bibinfo{year}{2005}).

\bibitem[{\citenamefont{Freericks and Monien}(1994)}]{Freericks94}
\bibinfo{author}{\bibfnamefont{J.~K.} \bibnamefont{Freericks}}
  \bibnamefont{and} \bibinfo{author}{\bibfnamefont{H.}~\bibnamefont{Monien}},
  \bibinfo{journal}{Europhys. Lett.} \textbf{\bibinfo{volume}{26}},
  \bibinfo{pages}{545} (\bibinfo{year}{1994}).

\bibitem[{\citenamefont{Freericks and Monien}(1996)}]{Freericks96}
\bibinfo{author}{\bibfnamefont{J.~K.} \bibnamefont{Freericks}}
  \bibnamefont{and} \bibinfo{author}{\bibfnamefont{H.}~\bibnamefont{Monien}},
  \bibinfo{journal}{Phys. Rev. B} \textbf{\bibinfo{volume}{53}},
  \bibinfo{pages}{2691} (\bibinfo{year}{1996}).

\bibitem[{\citenamefont{Elstner and Monien}(1999)}]{Elstner99}
\bibinfo{author}{\bibfnamefont{N.}~\bibnamefont{Elstner}} \bibnamefont{and}
  \bibinfo{author}{\bibfnamefont{H.}~\bibnamefont{Monien}},
  \bibinfo{journal}{Phys. Rev. B} \textbf{\bibinfo{volume}{59}},
  \bibinfo{pages}{12184} (\bibinfo{year}{1999}).

\bibitem[{\citenamefont{Buonsante and Vezzani}(2005)}]{Buonsante05}
\bibinfo{author}{\bibfnamefont{P.}~\bibnamefont{Buonsante}} \bibnamefont{and}
  \bibinfo{author}{\bibfnamefont{A.}~\bibnamefont{Vezzani}},
  \bibinfo{journal}{Phys. Rev. A} \textbf{\bibinfo{volume}{72}},
  \bibinfo{pages}{013614} (\bibinfo{year}{2005}).

\bibitem[{\citenamefont{Gros and Valent\'i}(1993)}]{Gros93}
\bibinfo{author}{\bibfnamefont{C.}~\bibnamefont{Gros}} \bibnamefont{and}
  \bibinfo{author}{\bibfnamefont{R.}~\bibnamefont{Valent\'i}},
  \bibinfo{journal}{Phys. Rev. B} \textbf{\bibinfo{volume}{48}},
  \bibinfo{pages}{418} (\bibinfo{year}{1993}).

\bibitem[{\citenamefont{S\'en\'echal et~al.}(2000)\citenamefont{S\'en\'echal,
  P\'erez, and Pioro-Ladri\`ere}}]{Senechal00}
\bibinfo{author}{\bibfnamefont{D.}~\bibnamefont{S\'en\'echal}},
  \bibinfo{author}{\bibfnamefont{D.}~\bibnamefont{P\'erez}}, \bibnamefont{and}
  \bibinfo{author}{\bibfnamefont{M.}~\bibnamefont{Pioro-Ladri\`ere}},
  \bibinfo{journal}{Phys. Rev. Lett.} \textbf{\bibinfo{volume}{84}},
  \bibinfo{pages}{522} (\bibinfo{year}{2000}).

\bibitem[{\citenamefont{S\'en\'echal et~al.}(2002)\citenamefont{S\'en\'echal,
  Perez, and Plouffe}}]{Senechal02}
\bibinfo{author}{\bibfnamefont{D.}~\bibnamefont{S\'en\'echal}},
  \bibinfo{author}{\bibfnamefont{D.}~\bibnamefont{Perez}}, \bibnamefont{and}
  \bibinfo{author}{\bibfnamefont{D.}~\bibnamefont{Plouffe}},
  \bibinfo{journal}{Phys. Rev. B} \textbf{\bibinfo{volume}{66}},
  \bibinfo{pages}{075129} (\bibinfo{year}{2002}).

\bibitem[{\citenamefont{Hubbard}(1963)}]{Hubbard63}
\bibinfo{author}{\bibfnamefont{J.}~\bibnamefont{Hubbard}},
  \bibinfo{journal}{Proc. Roy. Soc. A} \textbf{\bibinfo{volume}{276}},
  \bibinfo{pages}{238} (\bibinfo{year}{1963}).

\bibitem[{\citenamefont{Potthoff}(2003{\natexlab{a}})}]{Potthoff03a}
\bibinfo{author}{\bibfnamefont{M.}~\bibnamefont{Potthoff}},
  \bibinfo{journal}{Eur. Phys. J. B} \textbf{\bibinfo{volume}{32}},
  \bibinfo{pages}{429} (\bibinfo{year}{2003}{\natexlab{a}}).

\bibitem[{\citenamefont{Potthoff}(2003{\natexlab{b}})}]{Potthoff03b}
\bibinfo{author}{\bibfnamefont{M.}~\bibnamefont{Potthoff}},
  \bibinfo{journal}{Eur. Phys. J. B} \textbf{\bibinfo{volume}{36}},
  \bibinfo{pages}{335} (\bibinfo{year}{2003}{\natexlab{b}}).

\bibitem[{\citenamefont{Potthoff et~al.}(2003)\citenamefont{Potthoff, Aichhorn,
  and Dahnken}}]{Potthoff03c}
\bibinfo{author}{\bibfnamefont{M.}~\bibnamefont{Potthoff}},
  \bibinfo{author}{\bibfnamefont{M.}~\bibnamefont{Aichhorn}}, \bibnamefont{and}
  \bibinfo{author}{\bibfnamefont{C.}~\bibnamefont{Dahnken}},
  \bibinfo{journal}{Phys. Rev. Lett.} \textbf{\bibinfo{volume}{91}},
  \bibinfo{pages}{206402} (\bibinfo{year}{2003}).

\bibitem[{\citenamefont{Dahnken et~al.}(2004)\citenamefont{Dahnken, Aichhorn,
  Hanke, Arrigoni, and Potthoff}}]{Dahnken04}
\bibinfo{author}{\bibfnamefont{C.}~\bibnamefont{Dahnken}},
  \bibinfo{author}{\bibfnamefont{M.}~\bibnamefont{Aichhorn}},
  \bibinfo{author}{\bibfnamefont{W.}~\bibnamefont{Hanke}},
  \bibinfo{author}{\bibfnamefont{E.}~\bibnamefont{Arrigoni}}, \bibnamefont{and}
  \bibinfo{author}{\bibfnamefont{M.}~\bibnamefont{Potthoff}},
  \bibinfo{journal}{Phys. Rev. B} \textbf{\bibinfo{volume}{70}},
  \bibinfo{pages}{245110} (\bibinfo{year}{2004}).

\bibitem[{\citenamefont{Aichhorn et~al.}(2004)\citenamefont{Aichhorn, Evertz,
  {von der Linden}, and Potthoff}}]{Aichhorn04}
\bibinfo{author}{\bibfnamefont{M.}~\bibnamefont{Aichhorn}},
  \bibinfo{author}{\bibfnamefont{H.~G.} \bibnamefont{Evertz}},
  \bibinfo{author}{\bibfnamefont{W.}~\bibnamefont{{von der Linden}}},
  \bibnamefont{and} \bibinfo{author}{\bibfnamefont{M.}~\bibnamefont{Potthoff}},
  \bibinfo{journal}{Phys. Rev. B} \textbf{\bibinfo{volume}{70}},
  \bibinfo{pages}{235107} (\bibinfo{year}{2004}).

\bibitem[{\citenamefont{S\'en\'echal et~al.}(2005)\citenamefont{S\'en\'echal,
  Lavertu, Marois, and Tremblay}}]{Senechal05}
\bibinfo{author}{\bibfnamefont{D.}~\bibnamefont{S\'en\'echal}},
  \bibinfo{author}{\bibfnamefont{P.-L.} \bibnamefont{Lavertu}},
  \bibinfo{author}{\bibfnamefont{M.-A.} \bibnamefont{Marois}},
  \bibnamefont{and} \bibinfo{author}{\bibfnamefont{A.-M.~S.}
  \bibnamefont{Tremblay}}, \bibinfo{journal}{Phys. Rev. Lett.}
  \textbf{\bibinfo{volume}{94}}, \bibinfo{pages}{156404}
  (\bibinfo{year}{2005}).

\bibitem[{\citenamefont{Aichhorn and Arrigoni}()}]{Aichhorn05}
\bibinfo{author}{\bibfnamefont{M.}~\bibnamefont{Aichhorn}} \bibnamefont{and}
  \bibinfo{author}{\bibfnamefont{E.}~\bibnamefont{Arrigoni}},
  \bibinfo{note}{cond-mat/0502047, to appear in Europhys. Lett.}

\bibitem[{\citenamefont{Dahnken et~al.}()\citenamefont{Dahnken, Potthoff,
  Arrigoni, and Hanke}}]{Danhken05}
\bibinfo{author}{\bibfnamefont{C.}~\bibnamefont{Dahnken}},
  \bibinfo{author}{\bibfnamefont{M.}~\bibnamefont{Potthoff}},
  \bibinfo{author}{\bibfnamefont{E.}~\bibnamefont{Arrigoni}}, \bibnamefont{and}
  \bibinfo{author}{\bibfnamefont{W.}~\bibnamefont{Hanke}},
  \bibinfo{note}{cond-mat/0504618}.

\bibitem[{\citenamefont{Luttinger and Ward}(1960)}]{Luttinger60}
\bibinfo{author}{\bibfnamefont{J.~M.} \bibnamefont{Luttinger}}
  \bibnamefont{and} \bibinfo{author}{\bibfnamefont{J.~C.} \bibnamefont{Ward}},
  \bibinfo{journal}{Phys. Rev.} \textbf{\bibinfo{volume}{118}},
  \bibinfo{pages}{1417} (\bibinfo{year}{1960}).

\bibitem[{\citenamefont{{De Dominicis} and
  Martin}(1964{\natexlab{a}})}]{Dedominicis64a}
\bibinfo{author}{\bibfnamefont{C.}~\bibnamefont{{De Dominicis}}}
  \bibnamefont{and} \bibinfo{author}{\bibfnamefont{P.~C.}
  \bibnamefont{Martin}}, \bibinfo{journal}{J. Math. Phys.}
  \textbf{\bibinfo{volume}{5}}, \bibinfo{pages}{14}
  (\bibinfo{year}{1964}{\natexlab{a}}).

\bibitem[{\citenamefont{{De Dominicis} and
  Martin}(1964{\natexlab{b}})}]{Dedominicis64b}
\bibinfo{author}{\bibfnamefont{C.}~\bibnamefont{{De Dominicis}}}
  \bibnamefont{and} \bibinfo{author}{\bibfnamefont{P.~C.}
  \bibnamefont{Martin}}, \bibinfo{journal}{J. Math. Phys.}
  \textbf{\bibinfo{volume}{5}}, \bibinfo{pages}{31}
  (\bibinfo{year}{1964}{\natexlab{b}}).

\bibitem[{\citenamefont{Vilk and Tremblay}(1997)}]{Vilk97}
\bibinfo{author}{\bibfnamefont{Y.}~\bibnamefont{Vilk}} \bibnamefont{and}
  \bibinfo{author}{\bibfnamefont{A.-M.~S.} \bibnamefont{Tremblay}},
  \bibinfo{journal}{J. Phys. I} \textbf{\bibinfo{volume}{7}},
  \bibinfo{pages}{1309} (\bibinfo{year}{1997}).

\end{thebibliography}

\end{document}